\documentclass[prd,twocolumn,superscriptaddress,nofootinbib]{revtex4-2}
\usepackage{natbib}
\usepackage{graphicx}
\usepackage{bm}
\usepackage{amssymb}
\usepackage{color}
\usepackage{amsmath}
\usepackage{mathrsfs} 
\usepackage{amstext}
\usepackage[english]{babel}
\usepackage{latexsym}
\usepackage{array}             
\usepackage{multirow}         
\usepackage[usenames,dvipsnames]{xcolor}
\usepackage[colorlinks=true,citecolor=Blue,linkcolor=RubineRed,urlcolor=Blue]{hyperref}
\usepackage{tocloft}

\newcommand{\beq}{\begin{equation}}
\newcommand{\eeq}{\end{equation}}
\newcommand{\beqa}{\begin{eqnarray}}
\newcommand{\eeqa}{\end{eqnarray}}

\begin{document}

\title{Structural phase transition and its  critical dynamics from holography}
\author{Chuan-Yin Xia}
\affiliation{ Center for Gravitation and Cosmology, College of Physical 
and Technology, Yangzhou University, Yangzhou 225009, China}
\author{Hua-Bi Zeng}
\email{hbzeng@yzu.edu.cn}
\affiliation{ Center for Gravitation and Cosmology, College of Physical 
and Technology, Yangzhou University, Yangzhou 225009, China}
\author{Chiang-Mei Chen}
\affiliation{ Department of Physics, Center for High Energy and High Field Physics (CHiP),
National Central University, Chungli 32001, Taiwan}
\author{Adolfo del Campo\href{https://orcid.org/0000-0003-2219-2851}{\includegraphics[scale=0.05]{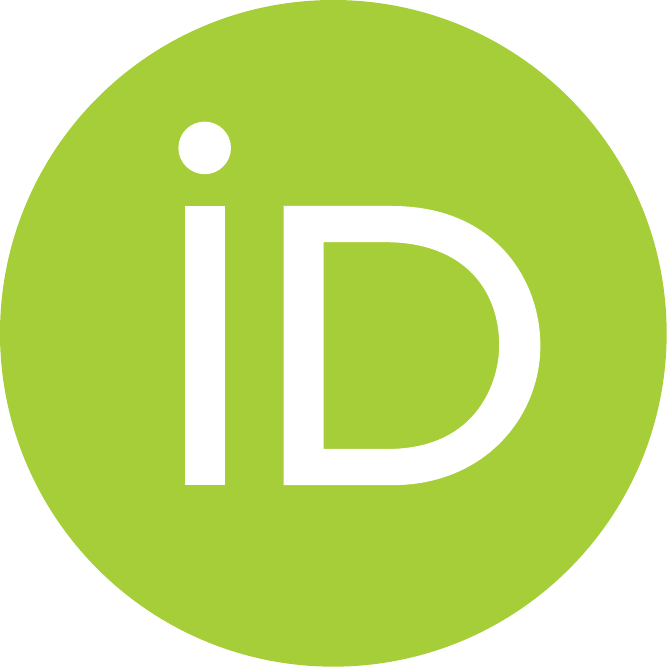}}}
\email{adolfo.delcampo@uni.lu}
\address{Department of Physics and Materials Science, University of Luxembourg,
L-1511 Luxembourg, Luxembourg}
\address{Donostia International Physics Center, E-20018 San Sebasti\'an, Spain}

\begin{abstract}
We introduce a gravitational lattice theory defined in an AdS$_3$ black hole background that provides a holographic dual description of the linear-to-zigzag structural phase transition, characterized by the spontaneous breaking of parity symmetry observed in, e.g., confined Coulomb crystals.
The transition from the high-symmetry linear phase to the broken-symmetry doubly-degenerate zigzag phase can be driven by quenching the coupling between adjacent sites through the critical point. An analysis of the equilibrium correlation length and relaxation time reveals mean-field critical exponents. We explore the nonequilibrium phase transition dynamics leading to kink formation. The kink density obeys universal scaling laws in the limit of slow quenches, described by the Kibble-Zurek mechanism (KZM), and at fast quenches, characterized by a universal breakdown of the KZM.

\end{abstract}

\maketitle

\section{Introduction}
The AdS/CFT correspondence provides an efficient and unique method to calculate the partition function of a strongly interacting field
theory from its dual gravitational theory, in one more spatial dimension~\cite{Maldacena1999,Gubser1998, Witten1998}.
The AdS/CFT correspondence has found successful applications in condensed matter theory (CMT),  under the umbrella of ``AdS/CMT''. These applications range from the universality of phase transitions to the exploration of exotic states of matter in the strongly coupled regime, in and out of equilibrium \cite{zaanen2015, ammon2015, hartnoll2018,Liu2020,Zaanen2019,zaanen2021}.
Here we address an essential paradigm in CMT that has not previously been captured in the scheme of AdS/CMT: the 
structural phase transition in crystalline solids~\cite{bruce1981, Cowley1980, Bruce1980, Gufan1982, Levanyuk1988, fujimoto2004, muller2012}. It involves a phase transition between different lattice configurations when the parameters in the external environment, such as temperature or pressure, change \cite{Saljebu1991}. 

The use of the linear Paul trap in trapped-ion physics has made it possible to observe, at low temperatures achieved by laser cooling, the arrangement of multiple ions into Coulomb crystals, and to explore different phases by adjusting the confining potential \cite{Thompson2015, Wieman1999, Leibfried2003}. Analogous experiments can be performed with confined colloids and dusty plasma \cite{Mansoori2014}. Specifically, structural phase transitions between different phases can be induced by axial compression, weakening the transverse confinement, or increasing the axial density (e.g., by cooling an ion plasma). In the limit of tight transverse confinement, Coulomb crystals form a linear chain. Weakening the transverse confinement induces a phase transition to a zigzag phase. Further weakening of the transverse confinement gives rise to higher dimensional structures. The associated phase transitions are generally of first order. An exception to this trend is offered by the structural phase transition between the linear chain and the doubly-degenerated zigzag phase \cite{Hasse1990,Schiffer1993,Dubin1993,Piacente2004,Retzker08}. The latter has been observed in various experiments ~\cite{Birkl1992,Raizen1992,Ejtemaee2013,Ulm2013,Pyka2013,Partner2015,Schneider2012}. It is a continuous phase transition resulting in the breaking of parity symmetry. The transverse width of the chain serves as an order parameter, being zero in the linear configuration and taking a finite value in the zigzag phase. The soft mode governing its growth in a critical quench is the transverse vibrational mode with the shortest wavelength~\cite{Piacente2004,Fishman2008,delCampo2010}. A lattice model~\cite{LagunaZurek1997,LagunaZurek1998,delCampo2010,DeChiara10,GomezRuiz2020,delCampo2022loc}  captures the phase transition dynamics, where there is a local order parameter $\phi_n$ defined on each lattice point, and an interaction potential  $\sum_n c\phi_n\phi_{n+1}$ 
is introduced to replace the standard second derivative term $\partial_x^2 \psi$ of the time-dependent Ginzburg-Landau equation. Thus, the critical dynamics,  and the formation of topological defects resulting from it,  can be described by the coupled Langevin  equations 
\begin{eqnarray}
&&\partial_t^2\phi_n+\eta\partial_t\phi_n+\partial_{\phi_n}
\sum_m[\lambda(t)\phi_m^2+\phi_m^4+c\phi_m\phi_{m+1}]
\nonumber\\
&& +\zeta(t)=0, \quad n=1,\dots,N, \label{eq1}
\end{eqnarray}
with constant friction $\eta>0$, and coupling $c>0$ favoring the zigzag ferromagnetic order. The real Gaussian white noise $\zeta(t)$ has zero mean and satisfies $\langle\zeta(t)\zeta(t+s)\rangle\propto\delta(s)$. The critical point of the phase transition is determined by the coupling constant as $\lambda_c=2c$.

The finite-time crossing of the phase transition results in zigzag domains of finite size with $\mathbb{Z}_2$ kinks at the interface between adjacent domains. This makes it an interesting test bed to probe the principles of nonequilibrium statistical mechanics. Across a continuous phase transition induced by a slow quench, the average domain size and the defect density are expected to scale with the quench time in which the transition is driven, following universal power laws predicted by the  Kibble-Zurek mechanism (KZM)~\cite{Kibble1976,Kibble1980,Zurek1985,Zurek1996,Polkovnikov2011,delCampo2014}. The test of the KZM across a structural phase transition was proposed in \cite{delCampo2010,DeChiara10} and has been realized using ion crystals ~\cite{Ejtemaee2013,Ulm2013,Pyka2013,DKZ13,Partner2015}. At fast quenches, by contrast, such scaling laws exhibit a universal breakdown and the domain size and defect density saturate to a value independent of the quench rate \cite{Zeng2023}. 

This work introduces a holographic model exhibiting a linear-to-zigzag phase transition. An analysis of the equilibrium critical properties reveals mean-field behavior. In particular, the power laws governing the divergence of the equilibrium correlation length and the relaxation time are consistent with critical exponents $\nu=1/2$ and $z=2$. We use this setting to explore the validity of the KZM across a structural phase transition for slow quenches and characterize the universal dynamics in the limit of fast quenches, when the KZM breaks downs. 

\section{Holographic version of one-dimensional lattice $\phi^4$ model}
To introduce a  gravity theory that accounts for the structural phase transition, we consider a ($d+1)$-dimensional anti–de Sitter (AdS$_{d+1}$) spacetime, the Reissner-Nordstrøm (RN) black hole background with $N$ discrete neutral scalar fields.
The RN black hole is a solution of the Einstein-Maxwell theory with negative cosmological constant $\Lambda = - d  (d - 1)/2 \ell^2$,
\begin{equation}
\mathcal{L} = R - 2 \Lambda + \alpha F_{\mu\nu} F^{\mu\nu}.
\end{equation}
We impose the following ansatz
\begin{equation}
ds^2 = \frac{\ell^2}{z^2} \left( - f(z) dt^2 + \frac{dz^2}{f(z)} + dx_{d-1}^2 \right), \quad A = A_t(z) dt.
\end{equation}
We further focus on the  $d = 2$ case to study a one-dimensional chain on the boundary, where
\begin{equation}
A_t = - \mu \ln z, \quad f = 1 - z^2 + \frac{\mu^2}{2} z^2 \ln z,
\end{equation}
with temperature
\begin{equation}
T = \frac1{4 \pi} \left( 2 - \frac{\mu^2}{2}\right).
\end{equation}
In the Eddington coordinate, $dt \to dt + dz/f$, the metric has the form
\begin{equation} \label{dsE}
ds^2 = \frac{\ell^2}{z^2} \left( -f(z) dt^2 - 2 dtdz + dx^2 \right).
\end{equation}

In the background of the RN black hole, the holographic theory of a structural
phase transition  involves $N$ real scalar fields $\Phi_n$ ($n = 1, \cdots, N$), governed by the Lagrangian
\begin{equation}
\mathcal{L}_{N} =\sum_{n=1}^{N} \frac{1}{2 \kappa \lambda} \left[ -\frac{1}{2} (\partial_z \Phi_n)^2 - V_M(\Phi_n) - V_c(\Phi_n, \Phi_{n+1}) \right]. \end{equation}
The value of $N$ fixes the number of lattice sites. One of the potentials entering the Lagrangian is given by the nonlinear Mexican hat potential  
\begin{equation}
V_M = \frac{1}{4 \ell^2} (\Phi_n^2 + m^2 \ell^2)^2 - \frac{m^4 \ell^2}{4},
\end{equation}
while the second one accounts for the   $z$-dependent coupling between neighboring sites
\begin{equation}
V_C = C z^2 \Phi_n \Phi_{n+1}.
\end{equation}
Since the coupling  vanishes at the boundary $z=0$, it 
will not change the boundary expansions of  the 
fields $\Phi_n$. As in the
lattice $\phi^4$ theory, the coupling 
potential can be treated as the substitute of the kinetic term $\partial_x\Phi\partial^x\Phi$.
In the background~Eq. \eqref{dsE}, we introduce  new variables $\psi_n = \Phi_n/z^{\frac{1}{2}}$ and choose a mass value $m^2=-3/4$.  Setting $\ell=1$, the equations of motion read
\begin{eqnarray}
&&2\partial_t\partial_z\psi_n=z^2(C(\psi_{n-1}+\psi_{n+1})-(f\partial^2_z +\partial_zf\partial_z)\psi_n)
\nonumber\\
&&- \frac{3-3f+2z \partial_z f}{4 }\psi_n
 +z\psi_n^3 , \quad n=1,\dots,N. \label{EOM}
\end{eqnarray}
The asymptotic expansions of the fields near the boundary are
\begin{equation}
\psi_n\big|_{z=0} =  A_n(t)+B_n(t) z .
\end{equation}
Standard quantization is adopted at the boundary, where $A_n$ can be regarded as the source of the operator in the boundary field theory, and $B_n$ can be regarded as the expected value of the scalar operator $\mathcal{O}$. Close the source of the operator, with $A_n$=0, one obtains a spontaneous symmetry-breaking state in this holographic setting.
We further notice that when all fields are decoupled by setting $C=0$, the Lagrangian of this discrete fields model reduces to the analog of the continuous field phase model in~\cite{Iqbal2010} with 
$d=3$. As detailed in the appendix \ref{AppA}, the model has a second-order phase transition  with mean-field
critical exponents,  and is driven by reducing the black hole temperature. The critical temperature decreases with the increase of mass, and when the mass squared $m^2$ is set to the BF upper bound $-d(d-1)/4$, it tends to zero temperature precisely, where a quantum phase transition occurs by reducing $m^2$.

\section{Equilibrium linear-to-zigzag phase transition and phase diagram}
\begin{figure}[t]
 \includegraphics[width=1.0\columnwidth,angle =0]{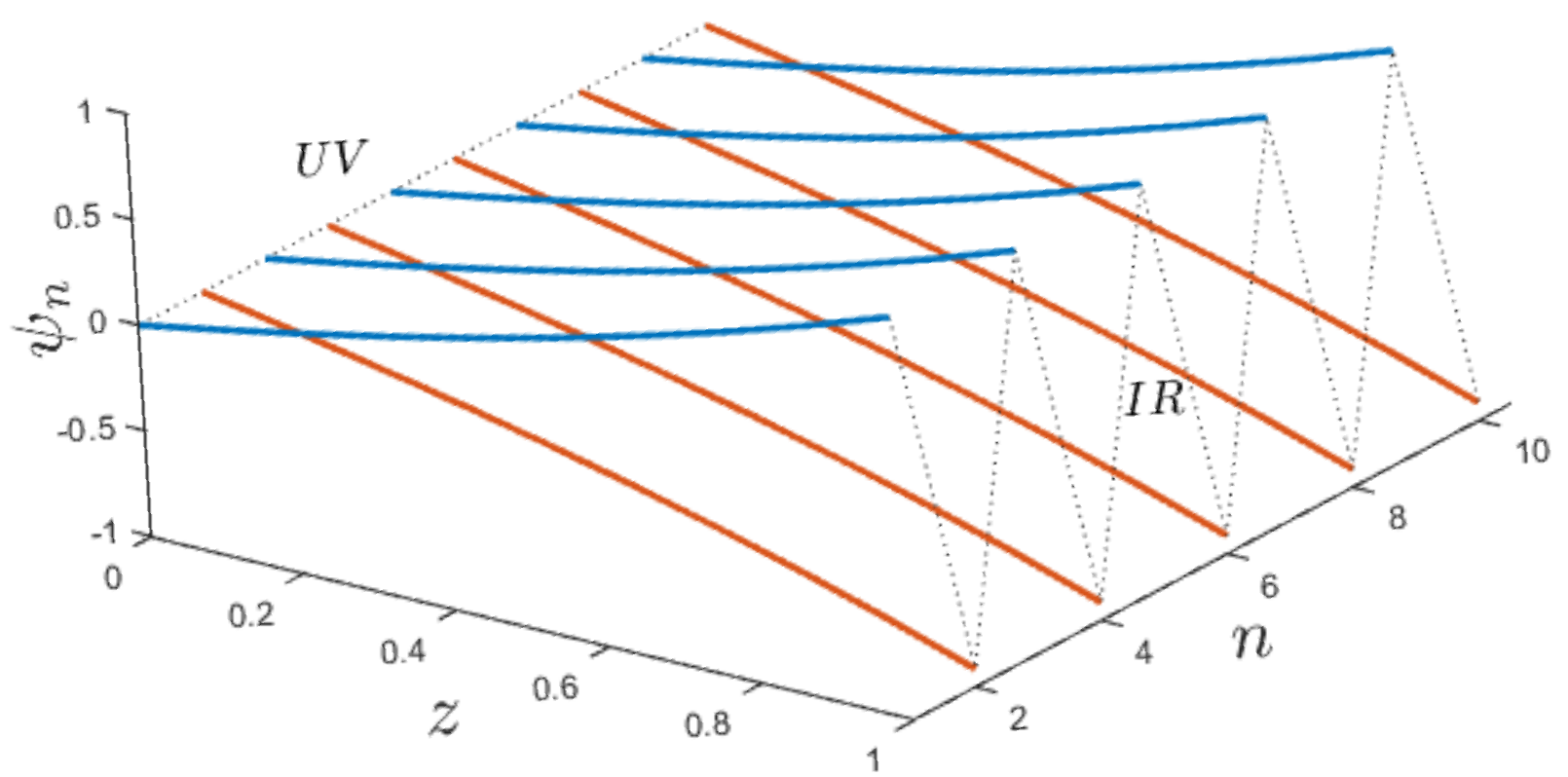}
\caption{Broken-symmetry configuration of the discrete fields $\psi_n$ from the horizon ($z=1$) to the boundary ($z=0$), where $N=10$, $C=1$, and $T=0.0796$. The zigzag structure on the horizon  is apparent in the field configuration for any finite $z>0$.
}\label{Fig1}
\end{figure}
\begin{figure}[t]
\includegraphics[trim=1.15cm 8.35cm 1.88cm 6.9cm, clip=true, scale=0.486, angle=0]{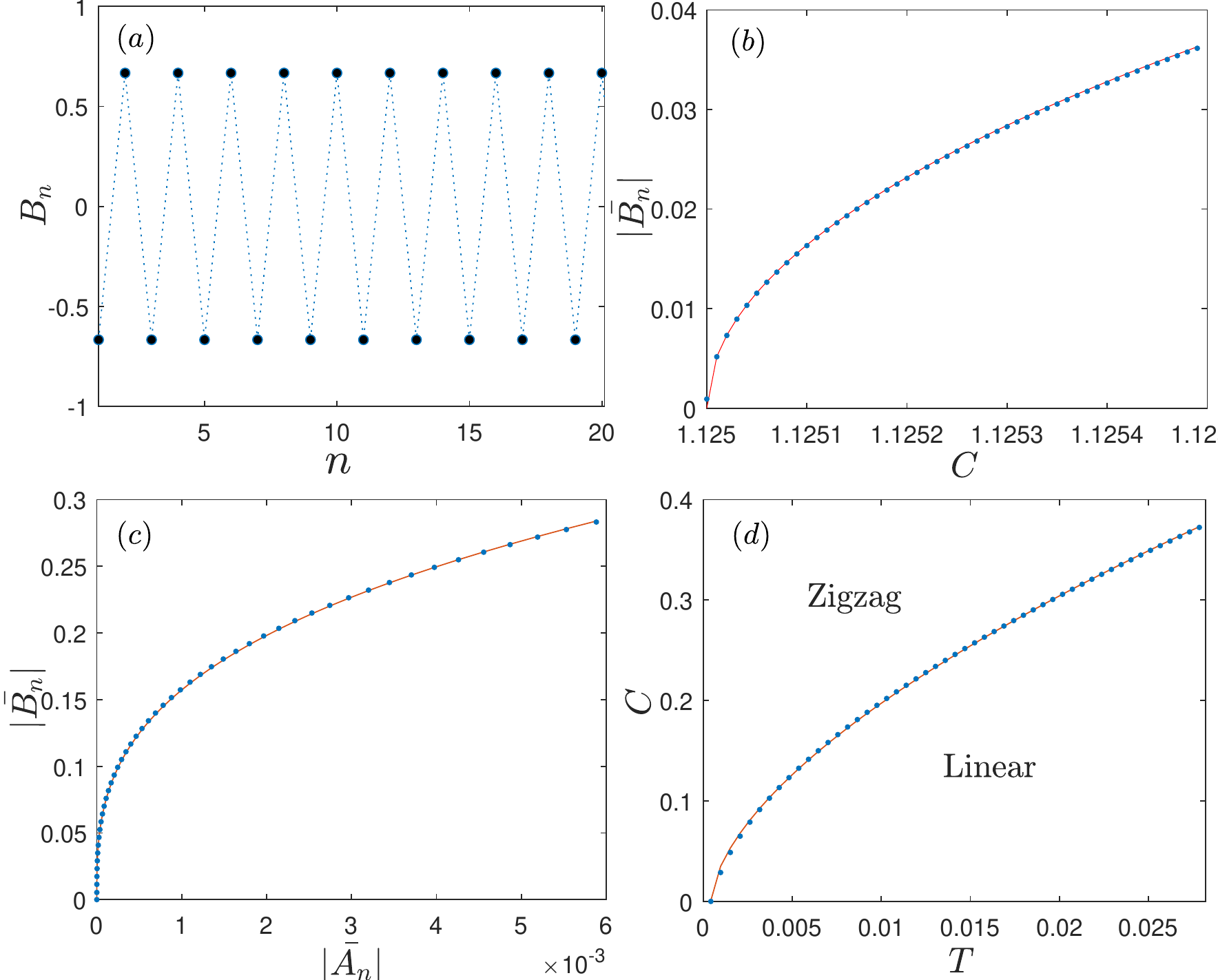}
\caption{Equilibrium properties of the holographic structural phase transition. (a) The local order parameter $B_n$ in one of the doubly-degenerate broken-symmetry zigzag phases as a function of the lattice coordinate  $n$, for $C=1$, $T=0.0796$, $N=90$. (b) Averaged absolute value of the order parameter $|\Bar{B}_n|$  as a function of the coupling $C$ at temperature  $T=0.1592$. The data is fitted to the power-law $(1.63\pm0.08)(C-C_c)^{0.500\pm 0.003}$, with $C_c=1.125$, shown as a red line. (c) Averaged  absolute value of the order parameter $|\Bar{B}_n|$ as a function of the averaged absolute value of the source $|\Bar{A}_n|$, at the critical point $C_c=1.125$ when $T=0.1592$. The red line is the corresponding  power-law fit    $(1.551\pm 0.004){|\Bar{A}_n|}^{1/(3.024\pm 0.004)}$. (d) Phase diagrams for the linear and zigzag phases. The curve embedded between the two phases is not linear, and fitted to the line $C=(3.28\pm 0.04)(T-4.27\times{10}^{-4})^{0.605\pm 0.003}$.
}\label{Fig2}
\end{figure}

Next, we  explore  the equilibrium phase transition of the holographic lattice model by setting the time derivative term in 
Eq.  \eqref{EOM} to zero and solving the reduced equation numerically by the Newton-Raphson iteration method. 
There is a critical coupling constant above which the $N$ scalar fields acquire a finite 
value, and a zigzag structure appears. 
A sample configuration  with 
$C=1$, $T=0.0796$ and $N=10$ is shown in Fig. \ref{Fig1}.  The $N$ fields are staggered in the $x$-direction as a positive value of the coupling coefficient is chosen, so that the system achieves the lowest free energy in the configuration $\psi_n(z)=|\psi_1(z)|(-1)^{n}$.
As shown in Fig. \ref{Fig2}(a), this leads to the so-called zigzag phase in which the order parameter has a saw-tooth profile in the condensed matter model. The coupling in the $x$-direction causes the signs of the local order parameter at adjacent lattice points to be opposite. 
The continuous phase transition of this model is induced by varying the coupling coefficient or the temperature. The phase diagram is shown in  Fig. \ref{Fig2}(d) as a function of the coupling coefficient $C$ and the temperature $T$. The critical line between the zigzag and linear phases shares a linear behavior as in the lattice $\phi^4$ model~\cite{GomezRuiz2020}. 
We fix $T=0.1592$, and the corresponding critical coupling is $C_c=1.125$.   In the proximity of the critical point, the  averaged absolute value of the order parameter $|\Bar{B}_n|$ is shown in Fig. \ref{Fig2}(b) as a function of the coupling constant $C$. The corresponding scaling as function of the  the  averaged  absolute value of the source $|\Bar{A}_n|$ is shown in Fig. \ref{Fig2}(c). The expected equilibrium scaling relations are $|\Bar{B}_n|\propto (C-C_c)^\beta$ and $|\Bar{B}_n|\propto |\Bar{A}_n|^{1/\delta}$.
The observed behavior is thus consistent with mean-field theory, where the critical exponents take values $\beta=1/2$ and $\delta=3$. 
According to the scaling and hyperscaling relations \cite{drouffe1989},
other critical exponents in the equilibrium phase transition can be computed:  $(\alpha,\beta,\gamma,\delta,\nu,\eta)=(0,\frac{1}{2},1,3,\frac{1}{2},0)$.
The value of the dynamical exponents $z=2$ can be extracted from a  quasi-normal modes analysis, detailed in the appendix \ref{AppQNN}.
This will be necessary to explore the universal critical dynamics, specifically,  the validity of the KZM scaling relations in the holographic setting, to which we next turn.

\section{Universal critical dynamics and defect formation: the Kibble-Zurek mechanism and its breakdown}

\begin{figure}[t]
\includegraphics[trim=1.6cm 7.4cm 1.9cm 7.8cm, clip=true,scale=0.496, angle=0]{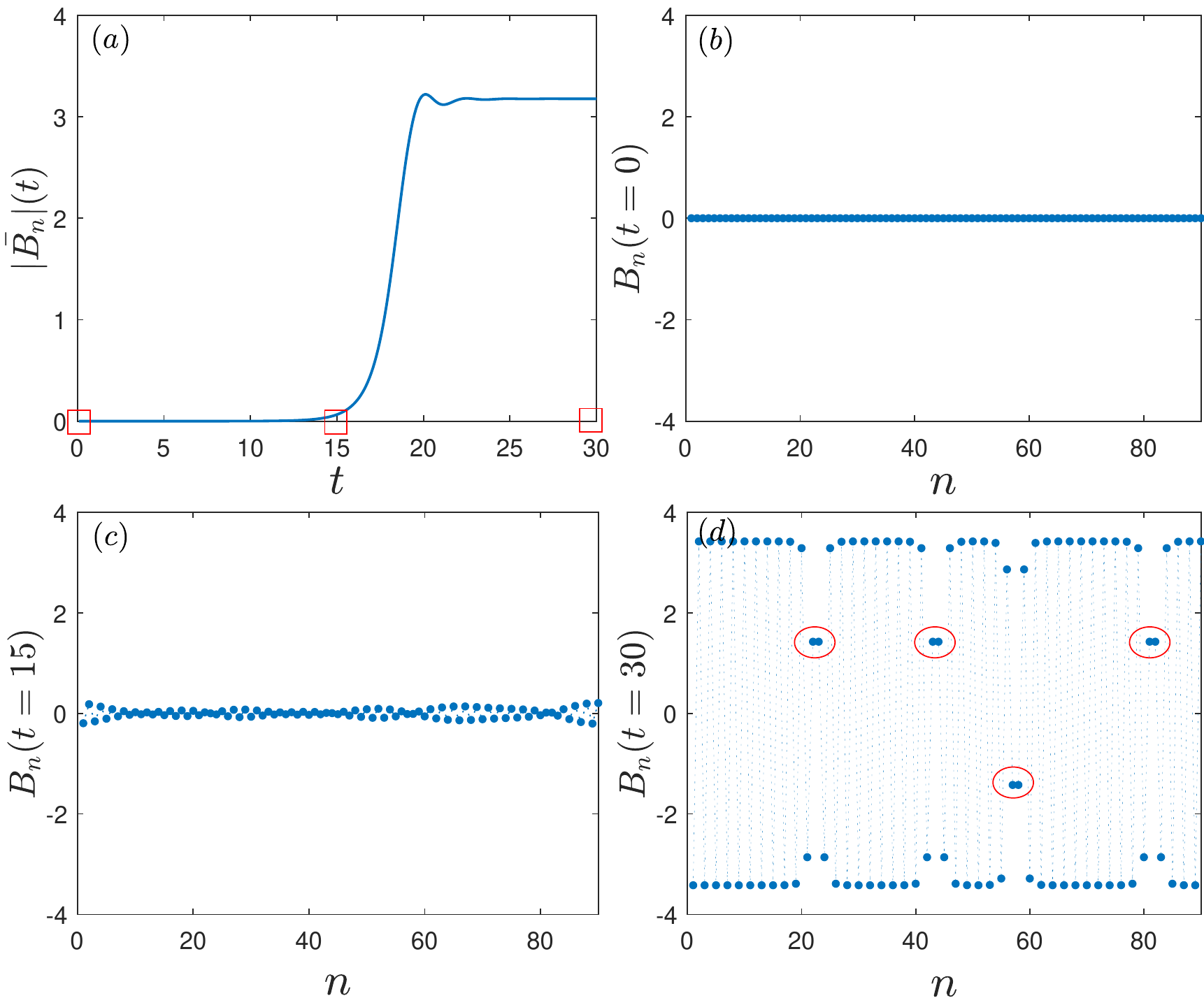}
\caption{Dynamics of the holographic structural phase transition. (a) Growth of the average order parameter as a function of time. Panels (b)-(c) correspond to snapshots of the spatial dependence of the local order parameter at different stages of the phase transition marked by red squares in panel (a), i.e.,  (b) $t=0$, (c) $t=15$  and (d) $t=30$ ($N=90$). While the order parameter vanishes in the high symmetry phase (b), it starts to acquire a finite value as the transition is crossed (b). (d) Eventually, the order parameter stabilizes with well-defined topological defects, known as $\mathbb{Z}_2$-kinks, marked by red circles at the interface of adjacent zigzag domains.}\label{Fig3}
\end{figure}

With the knowledge of the equilibrium properties, we next tackle the nonequilibrium phase transition by 
linearly quenching the coupling constant from   the
critical coupling value $C_c$ to the final coupling $C_f$ in a finite quench time $\tau_Q$. This protocol corresponds to ``half'' a quench, and is motivated by the fact that spontaneous symmetry breaking is governed by the dynamics after the critical point \cite{Antunes06}. The fourth-order Runge-Kutta method is used to numerically solve  Eq.  \eqref{EOM}.
We consider the linearized quench function 
\begin{equation}
 C(t)=C_c(1+t/\tau_Q).
\end{equation}
The time $t_f$  at which the modulated coupling reaches its final value, $C(t_f)=C_f$, will play a key role in what follows and is given by
\begin{equation}
t_f=\tau_Q (C_f-C_c)/C_c.\label{endingtime}
\end{equation}
The values $C_f=4$, $\tau_Q=5$, and $N=90$ are chosen in the simulation.
At the initial time, the system is in the linear phase. To seed symmetry breaking, random fluctuations in the fields in the bulk are used,  with a vanishing statistical average  $\langle b_n(t) \rangle=0$,  and two-point correlations $\langle b_{n}(t) b_{m}(t')\rangle\propto\delta_{nm}\delta(t-t')$.
The dynamics of symmetry breaking in this setting is shown in Fig. \ref{Fig3}, which shows the formation process of topological defects. 
During the transition,  the average value of the order parameter evolves with time,  increases monotonically early on. Subsequently, its time dependence exhibits oscillations as the saturation value in the broken-symmetry phase is approached, see Fig. \ref{Fig3}(a). Panels Fig. \ref{Fig3}(b)-(d) show the spatial dependence of the order parameter at different stages of the phase transition, displaying the evolution from the high-symmetry linear phase to an ordered zigzag structure. In \ref{Fig3}(b), the order parameter vanishes. The oscillation of  $|B_n|$ in \ref{Fig3}(c), before reaching the value associated with the zigzag phase,  reflects the dynamic instability of the one-dimensional configuration after crossing the critical point. With the subsequent growth of the order parameter, 
the completion of the transition results in different zigzag domains shown in \ref{Fig3}(d).
Kinks are identified by configurations in which the order parameter at adjacent sites has the same sign (marked by red ellipses). The formation of these $\mathbb{Z}_2$-kinks is predicted by the KZM. We note that the absolute value of the order parameters $|B|$ near the kinks (about  3.2) is smaller than that of regions of the zigzag domains (about 3.6). Thus, the average absolute value of the nonequilibrium configurations of the order parameter resulting from the critical dynamics is smaller than that in defect-free zigzag phase, 
due to the presence of kinks. 

\begin{figure}[t]
\includegraphics[width=1.0\columnwidth,angle =0]{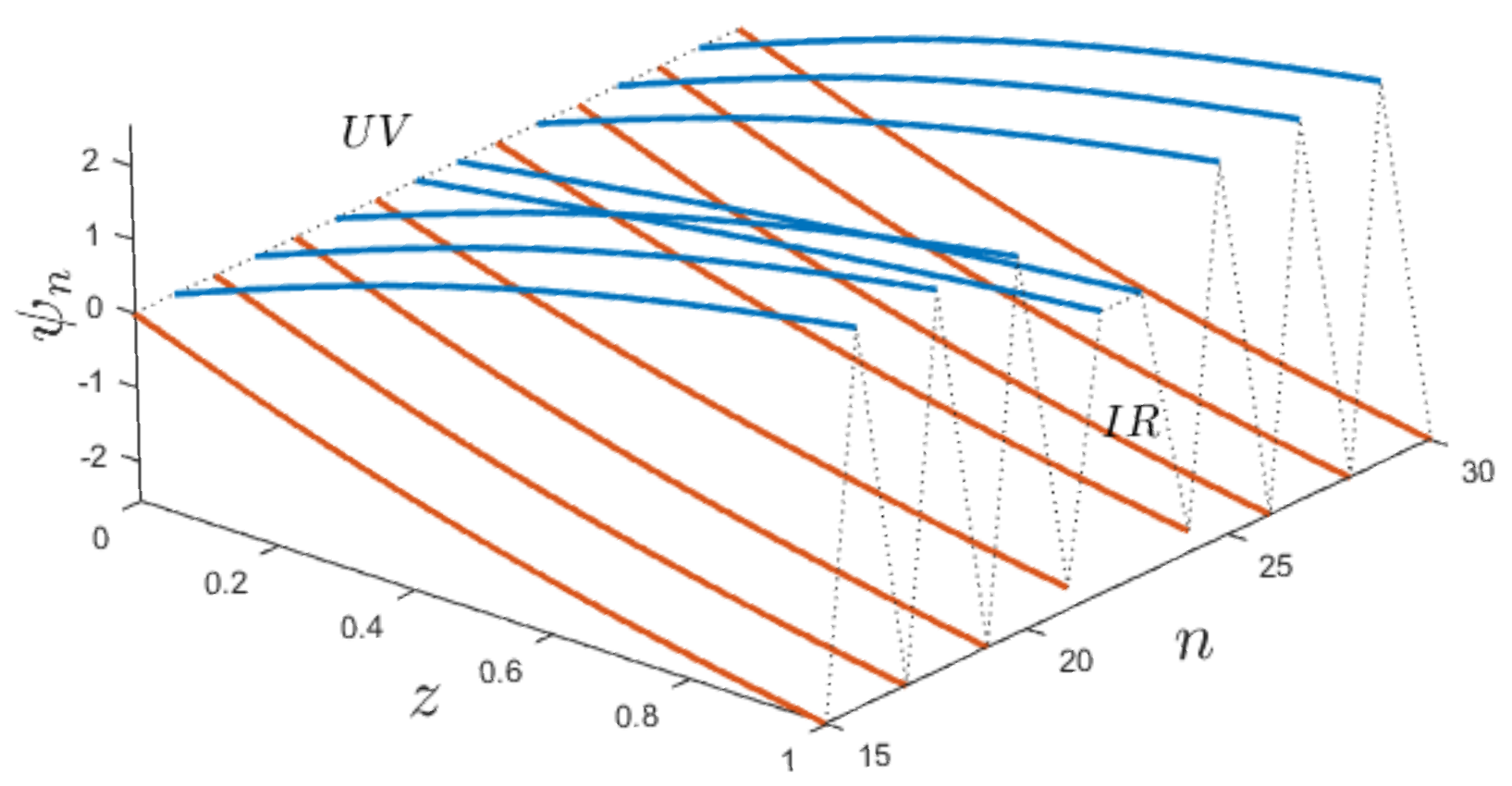}
\caption{Nonequilibrium configuration of the discrete fields $\psi_n$ in the zigzag phase, supporting  a kink from the horizon ($z=1$) to the boundary ($z=0$). This configuration is a holographic projection of the realization in Fig. \ref{Fig3} (d), with $C=4$ and $T=0.1592$, and is to be contrasted with the kink-free case shown in Fig. \ref{Fig1}. The effect of the kink is apparent for any finite $z>0$. 
}\label{Fig4kink}
\end{figure}
KZM is one of the few universal paradigms in nonequilibrium statistical mechanics and describes the critical dynamics across phases with different symmetries \cite{Kibble1976,Kibble1980,Zurek1985,Zurek1996,Polkovnikov2011,delCampo2014}. It exploits the equilibrium power-law divergence of the relaxation time $\tau=\tau_{0}|\epsilon|^{-z\nu}$ and the correlation length $\xi=\xi_{0}|\epsilon|^{-\nu}$ as a function of the parameter distance to the critical point $\epsilon=(C-C_c)$. Note that these scaling relations are governed by the correlation-length critical exponent $\nu$ and the dynamic critical exponent $z$. KZM predicts that in a system driven in a finite quench time $\tau_Q$, by varying $\epsilon(t)=t/\tau_Q$, the effective relaxation time is given by the freeze-out time $\hat{t}$. The latter is obtained by matching the instantaneous equilibrium relaxation time $\tau(t)=\tau_{0}|t/\tau_Q|^{-z\nu}$ to the time $t$ elapsed after crossing the phase transition. The freeze-out time thus scales as a power-law of the quench time $\tau_Q$, i.e., 
\begin{equation}
\hat{t}=(\tau_0 \tau_{Q}^{z\nu})^{\frac{1}{1+z\nu}}.\label{tfreeze1}
\end{equation}
The associated nonequilibrium correlation length  reads
\begin{equation}
\hat{\xi}=\xi_{0}|\epsilon(\hat{t})|^{-\nu}=\xi_0(\tau_{Q}/\tau_0)^{\frac{ \nu}{1+z\nu}},\label{xitf}
\end{equation}
with a power-law growth as the quench time is increased.
As shown in Fig. \ref{Fig3}, during the evolution, the holographic model is partitioned into several domains with average size $\hat{\xi}$. This leads to the formation of  kinks localized at the interface between adjacent domains. The nonequilibrium configuration of the discrete fields $\psi_n(z)$ in the presence of a kink in the zigzag phase is shown in Fig. \ref{Fig4kink} along the holographic direction. The presence of the kink is noticeable in $\psi_n(z)$ for any $z>0.$
The KZM estimates the average number of kinks  $n$ according to
\begin{equation}
n  \propto \frac{N}{\hat{\xi}} \propto\tau_{Q}^{-\frac{ \nu}{1+z\nu}},\label{kzlaw}
\end{equation}
which scales  universally with $\tau_Q$. The exponent in the power law (\ref{kzlaw}) is determined by the equilibrium critical exponents $\nu$ and $z$. The beauty of this KZM prediction relies on the fact that equilibrium critical scaling theory is used to describe  the nonequilibrium behavior, which inherits a universal scaling.

\begin{figure}[t]
\includegraphics[trim=3.9cm 12.4cm 4.11cm 12.6cm, clip=true, scale=0.4, angle=0]{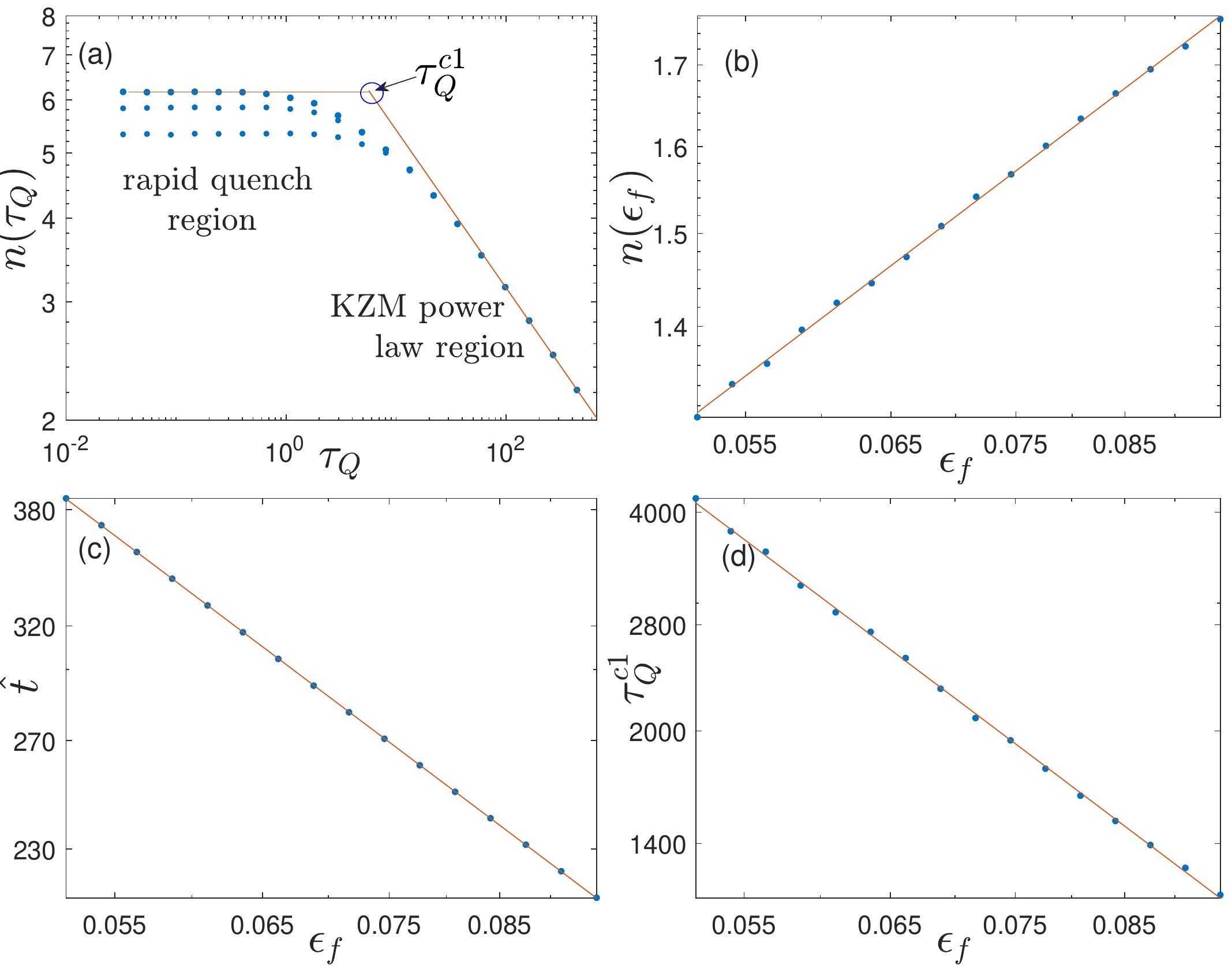}
\caption{Universal scaling laws at fast and slow quenches in a  log-log representation.
(a) The mean number $n$ of kinks as function of the quench time $\tau_Q$ ($N=100$). The behavior at slow quenches exhibits a universal  power-law scaling with the quench time, in agreement with KZM. As the quench time is reduced,  there is a crossover to a plateau region with a value of the kink average number that is independent of the quench time. The crossover occurs for $\tau_Q$ shorter than $\tau_Q^{c1}$. For rapid quenches, panels (b)-(d) show that the mean number $n$ of kinks, the freeze-out time $\hat{\xi}$, and the critical quenching time $\tau_Q^{c1}$ scale universally as a function of the quench depth $\epsilon_{f}$. The corresponding data is fitted by the following power-laws  $n=(5.59\pm 0.09)(C_f-C_c)^{0.491\pm 0.007}$, $\hat{t}=(20.96\pm 0.09)(C_f-C_c)^{-0.986\pm 0.001}$, and $\tau_{Q}^{c1}=(8.2\pm 0.7)(C_f-C_c)^{-2.11\pm 0.04}$.
}\label{Fig4}
\end{figure}

To explore its validity in the holographic linear-to-zigzag structural phase transition, Fig. \ref{Fig4} reports in a log-log scale the average number of topological defects as a function of quench time $\tau_Q$. Each sampling point is the average of $10^5$ realizations. We set the temperature $T=0.1592$ and the critical coupling constant $C_c=1.125$.
In panel \ref{Fig4}(a), from up to bottom, curves for the average number of kinks are shown for three different values of the end coupling constants $C_f=2, 3, 4$. The right sides of the three curves all meet on a common line with a negative slope,  indicative of universal nonequilibrium behavior, and  fitted to $n=(9.3\pm 0.4)\tau_Q^{-0.235\pm 0.08}$. This is in reasonable agreement with the universal KZM power-law scaling in Eq.  \eqref{kzlaw}, with a power-law exponent $\nu/(1+z\nu)=1/4$ using the verified values of the equilibrium critical exponents $\nu=1/2$ and $z=2$,  consistent with the calculation of quasi-normal modes.
The proximity of the numerical value of the fitted power-law exponent to the  KZM  prediction supports the validity of the mean-field description and the validity of the KZM in holography. These results are also consistent with the results of the Langevin equation for the zigzag phase transition in ion chains \cite{delCampo2010,DeChiara10} and the lattice $\phi^4$ model~\cite{LagunaZurek1997,LagunaZurek1998,GomezRuiz2020}. We further note that in the KZM scaling regime, the value of $t_f$ plays no significant role, as the breaking of symmetry and stabilization of topological defects happens in a time scale $t\sim\hat{t}$, that is small with respect to $t_f$.

The validity of the KZM is generally restricted to slow quenches, as long as they are far from the onset of adiabatic dynamics (expected when the domain size is comparable to the system size, $\hat{\xi}\sim N$). Indeed, in quantum systems, the KZM scaling laws have been derived using adiabatic perturbation theory \cite{Polkovnikov05,Dziarmaga05,DeGrandi10}. It is known that KZM scaling laws break down at moderate and fast quenches, due to a saturation of the defect density. One may thus expect that signatures of universality in the nonequilibrium dynamics are restricted to the limit of slow quenches. Contrary to this expectation, recent works have shown that the critical dynamics for moderate and fast quenches also admits a universal description, in holographic systems \cite{Chesler:2014gya}, as well as in classical and quantum systems \cite{Zeng2023}. This is a significant advance, as such conditions are of relevance to many scenarios, away from the large $\tau_Q$ limit.

Fig. \ref{Fig4}(a) reports the defect density generated by crossing the critical point in a wide range of quenching times. In particular, the left region of the curve, corresponding to rapid quenching, shows a plateau and indicates the breakdown of the KZM power-law as a function of the quench time $\tau_Q$. This plateau arises for different values of the end coupling constants $C_f$. 
This saturation of the defect number at fast quenches is consistent with observations in  experiments~\cite{Partner2015,Ulm2013,Pyka2013} and theoretical studies~\cite{Zeng2021,Li2020,Xia2020,Li2021,delCampo2021,GomezRuiz2020,delCampo2010,Ejtemaee2013}. 
In what follows, we not only elucidate the mechanism of the existence of the plateau region in the rapidly quenched region, but also deduce its value (the average number of defects), and characterize the crossover between the plateau and the KZM power-law  \cite{Zeng2023}. For rapid quenching, the end time $t_f$ is smaller than the relaxation time, that is
\begin{equation}
t_f\leq \tau(t_f)=\tau(C_f), \label{rapidconditon}
\end{equation}
so the freezing will not occur when $0<t<t_f$. As the time of evolution goes by and  $t$ increases, it eventually matches the minimum relaxation time $\tau(C_f)$. Thus, the freeze-out time  in a rapid quench reads
\begin{equation}
\hat{t}=\tau(C_f) \propto \epsilon_{f}^{-\nu z}. \label{rapidtf}
\end{equation}
In addition, the correlation length is also determined by the final coupling value $C_f$ at the freeze-out time, and the average number of defects, 
\begin{equation}
n\propto \frac{1}{\xi(C_f)} \propto \epsilon_{f}^\nu. \label{rapidn}
\end{equation}
To sum up, quench protocols that satisfy the constraints of Eq. \eqref{rapidconditon} can be called rapid. They lead to a nonequilibrium configuration of the system that is solely determined by the final value of the coupling constant $C_f$ and the equilibrium critical exponents. As a result, the defect density Eq. \eqref{rapidn} forms a plateau in the rapid quenching region, independent of the quench time. 
Furthermore, by matching the quench end time $t_f$ and the relaxation time $\tau(C_f)$, the crossover quenching time $\tau_{Q}^{c1}$ can be obtained. Substituting Eq. \eqref{xitf} and Eq. \eqref{endingtime} into Eq. \eqref{rapidconditon}, i.e., 
\begin{equation}
\tau_{Q}^{c1}(C_{f}-C_{c})/C_{c}=\tau_{0}(C_{f}-C_{c})^{- \nu z},
\end{equation}
yields 
\begin{equation}
\tau_{Q}^{c1}\propto\epsilon_{f}^{-(\nu z+1)}.\label{tauQc1}
\end{equation}
We accurately verified the power law of  Eqs. \eqref{rapidtf}, \eqref{rapidn}, and \eqref{tauQc1} in the rapid quench region in Fig. \ref{Fig4}(b)-(d). These results establish the universality of critical dynamics in rapid quenching in a holographic framework.

\section{Summary}
We have introduced a  gravitational theory defined
in the background of an AdS$_3$ RN black hole
as a holographic dual description of the linear-to-zigzag structural
phase transition, exhibited in confined interacting particles such as trapped ions and colloids. This theory provides a natural test bed to study holographic structural phase transitions in and out of equilibrium. The theory includes
$N$ scalar fields living on 
a chain with a positive coupling between the nearest lattice sites. Increasing the coupling constant or reducing the black hole temperature, the continuous linear-to-zigzag phase transition with Ginzburg-Landau universality is found. 

We have shown that the dynamics induced by a finite-time quench  across
the structural phase transition results in the formation of kinks, the density of which scales universally with the quench rate,  as described by the Kibble-Zurek mechanism, for slow quenches. This behavior is analogous to that previously reported in classical models of driven Coulomb crystals in the overdamped regime \cite{delCampo2010,DeChiara10}. Our study further reveals a universal breakdown of the KZM in the fast quench limit, in which the defect density forms a plateau independent of the quench time. The characteristic quench time associated with the crossover between the KZM and plateau regimes, as well as the average defect number at the plateau,  exhibit a universal power-law scaling as a function of the final value of the control parameter driving the transition and the equilibrium critical exponents. 

Our results thus established the validity of the universal scaling laws for arbitrary quenches across a structural phase transition, from the slow \cite{delCampo2014} to the fast limit \cite{Zeng2023},  in a holographic setting. We expect our findings to hold in conventional structural phase transitions in condensed matter systems, such as confined Coulomb crystals, colloids, and dusty plasma. In this context, our findings motivate a new generation of experimental studies aimed at probing the universality of defect formation in scenarios of spontaneous symmetry breaking induced by fast quenches.

Our model may also inspire further progress on the universality of critical dynamics in holographic systems. Interesting prospects include the identification of signatures intrinsic to holography, strong coupling, and the quest for behavior beyond mean-field \cite{Chesler:2014gya,Sonner:2014tca,Liu2020}, as well as the treatment of quantum fluctuations \cite{Retzker08,Landa10,Shimshoni11,Cormick12,Silvi16} in the holographic setting. Variants of the model put forward may accommodate for other continuous structural phase transitions, such as that between the linear configuration and a helix \cite{Nigmatullin2016}, as well as first-order structural phase transitions \cite{Mansoori2014}. In addition, our results make possible the study of signatures of universality beyond the mean defect density predicted by the KZM, e.g., in the full distribution of the number of defects \cite{delcampo18,Cui19,GomezRuiz2020,Bando20,Mayo21,delCampo2021,Subires22,GomezRuiz22,King2022} and their spatial statistics \cite{delCampo2022loc}. 
\acknowledgements

We thank Jan Zaanen and Li Li for their valuable comments.
This work is supported by the National Natural Science Foundation of China (under Grant No. 12275233) and the Postgraduate Research $\&$ Practice Innovation Program of Jiangsu Province (KYCX22$\_$3450).

\appendix

\section{Phase diagram of the holographic model without coupling in AdS$_3$}\label{AppA}

\begin{figure}[t]
\includegraphics[trim=3.82cm 8.8cm 4.3cm 9.5cm, clip=true, scale=0.65, angle=0]{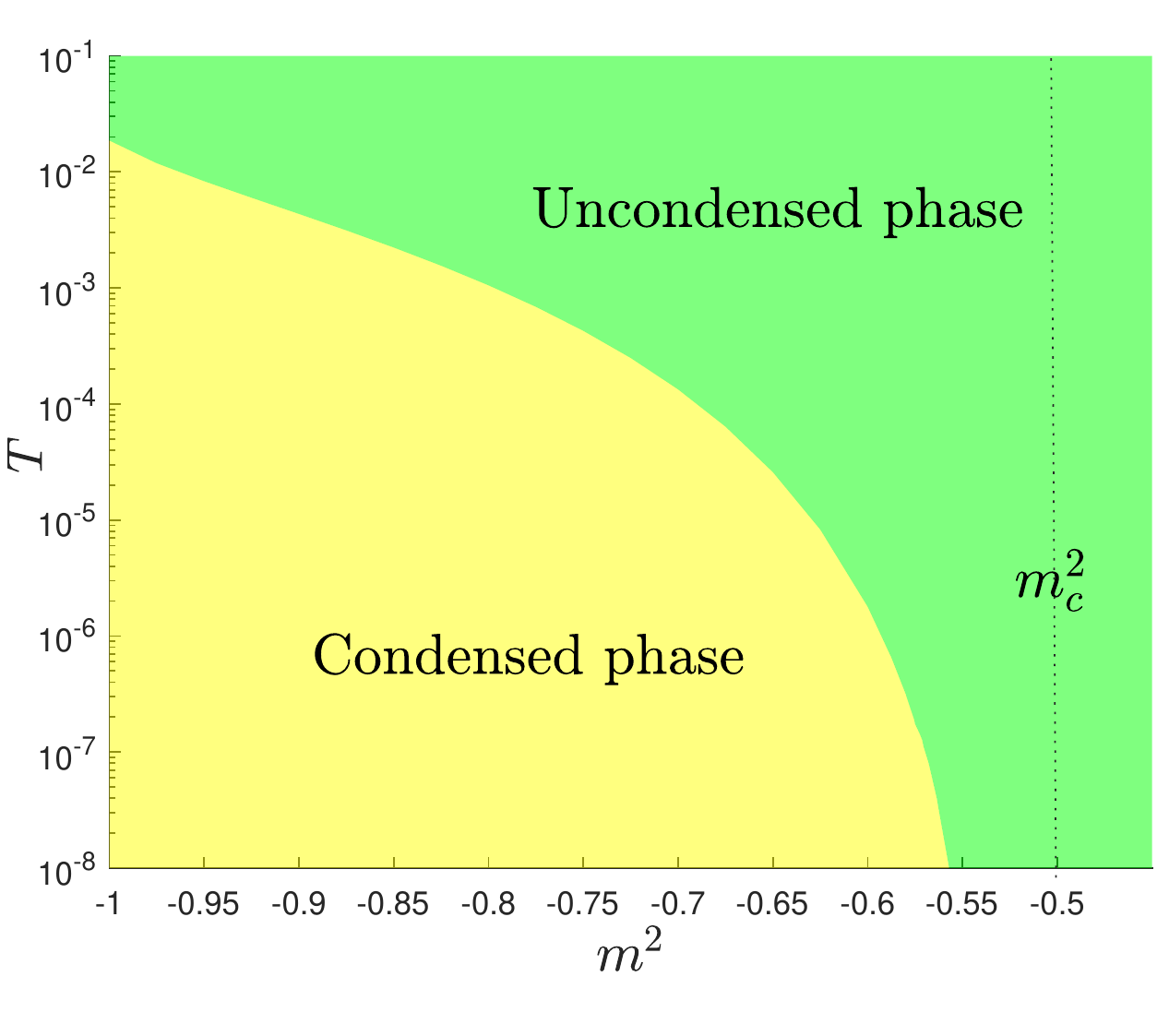}
\caption{Phase diagram of the holographic model when $C=0$. The critical temperature $T_c$ tends to zero when $m^2$ approaches $m_c^2$, leading to a quantum critical point.
}\label{Fig5}
\end{figure}
\begin{figure}[t]
 \includegraphics[width=1.0\columnwidth,angle =0]{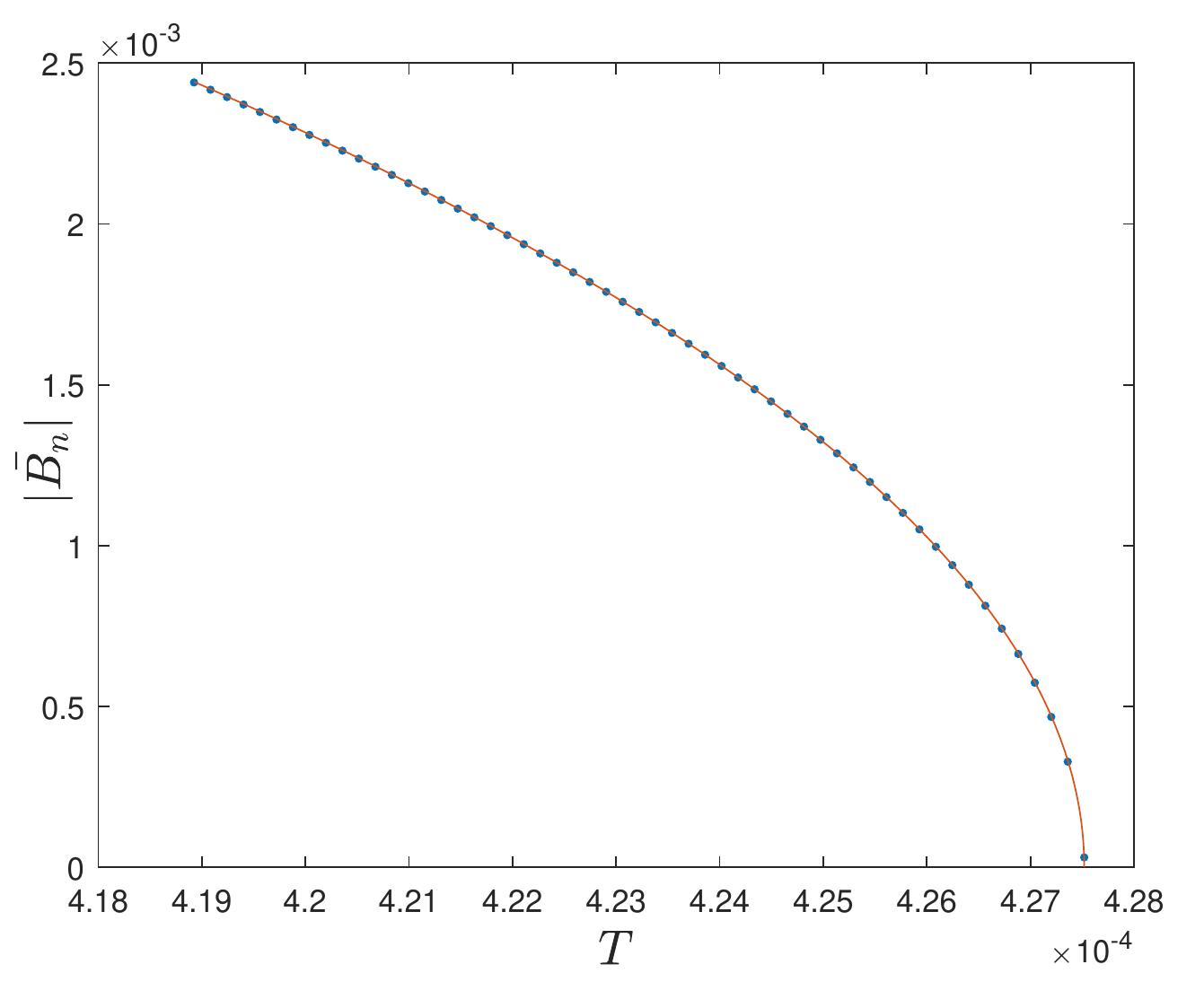}
\caption{Temperature-driven phase transition without coupling: dependence of the averaged absolute value of  the order parameter $|\Bar{B}_n|$ as a function of the temperature. The blue dots represent numerical results, while the solid red line shows the fit to the data $(0.833\pm 0.009)(T_c-T)^{0.49\pm 0.01}$, where $T_c=4.275 \times 10^{-4}$.
}\label{Fig6}
\end{figure}
The choice $N=1$ or  coupling $C=0$   leads to a vanishing $V_C$, and the Lagrangian of this  model  with discrete fields reduces to the continuous field phase model in Ref.~\cite{Iqbal2010},
 \begin{equation}
\mathcal{L} =\frac{1}{2 \kappa \lambda} \left[ -\frac{1}{2} (\partial_z \phi)^2 -  \frac{1}{4 \ell^2} (\phi^2 + m^2 \ell^2)^2 + \frac{m^4 \ell^2}{4} ) \right], \nonumber
\end{equation}
 but defined in the 
$d=3$ case.
For equilibrium phase transitions, the scalar field can be seen as a function of the holographic coordinate $z$. The equation of motion in the Eddington coordinate is
\begin{equation}
    -z \left(z f' \phi '+f \left(z \phi ''-\phi '\right)\right)+m^2 \phi +\phi^3= 0.
\end{equation}
The asymptotic expansions of the field near the boundary is
\begin{equation}
\phi\big|_{z=0} = \phi^+ z^{\triangle^+} + \phi^- z^{\triangle^-},
\end{equation}
where
\begin{equation}
\triangle^\pm = \frac{2 \pm \sqrt{4 + 4 m^2}}{2}.
\end{equation}
 We adopt the standard quantization, where $\phi^-$ is regarded as the source of the operator in boundary field theory, which is always set to 
 be vanishing to study the spontaneous 
 symmetry broken state. Then, $\phi^+$ is regarded as the expected value of the
scalar operator $\mathcal{O}$, the order parameter in the broken-symmetry  phase. 

As in the AdS$_4$ case, a continuous phase transition occurs by reducing the black hole temperature. We draw the phase diagram of the equilibrium phase transition in Fig. \ref{Fig5} by solving the equation for $\phi$. The phase diagram shows that $T_c$ tends to zero when $m^2$ near the BF-bound $m_c^2=-1/2$, leading to a quantum critical point.
By setting the mass to $m^2=-3/4$  as an example, we plot the condensation $\mathcal{O}$ as a function of temperature $T$ in Fig.  \ref{Fig6}. The curve can be fitted by a power-law $0.833(T_c-T)^{1/2}$, which reveals a critical exponent of the mean-field type, $\beta=1/2$.
These results are similar to the  AdS$_4$ model in Ref.~ \cite{Iqbal2010}. This is expected since the holographic model is defined at the classical level, where the boundary field theory is in the large $N$ limit: Fluctuations are then  suppressed, justifying the appearance of the 
mean-field  universality  in systems in both two and three spatial dimensions.

\section{Quasi-Normal Modes (QNMs) and the dynamic critical exponent $z$}\label{AppQNN}

\begin{figure}[t]
 \includegraphics[width=1.0\columnwidth,angle =0]{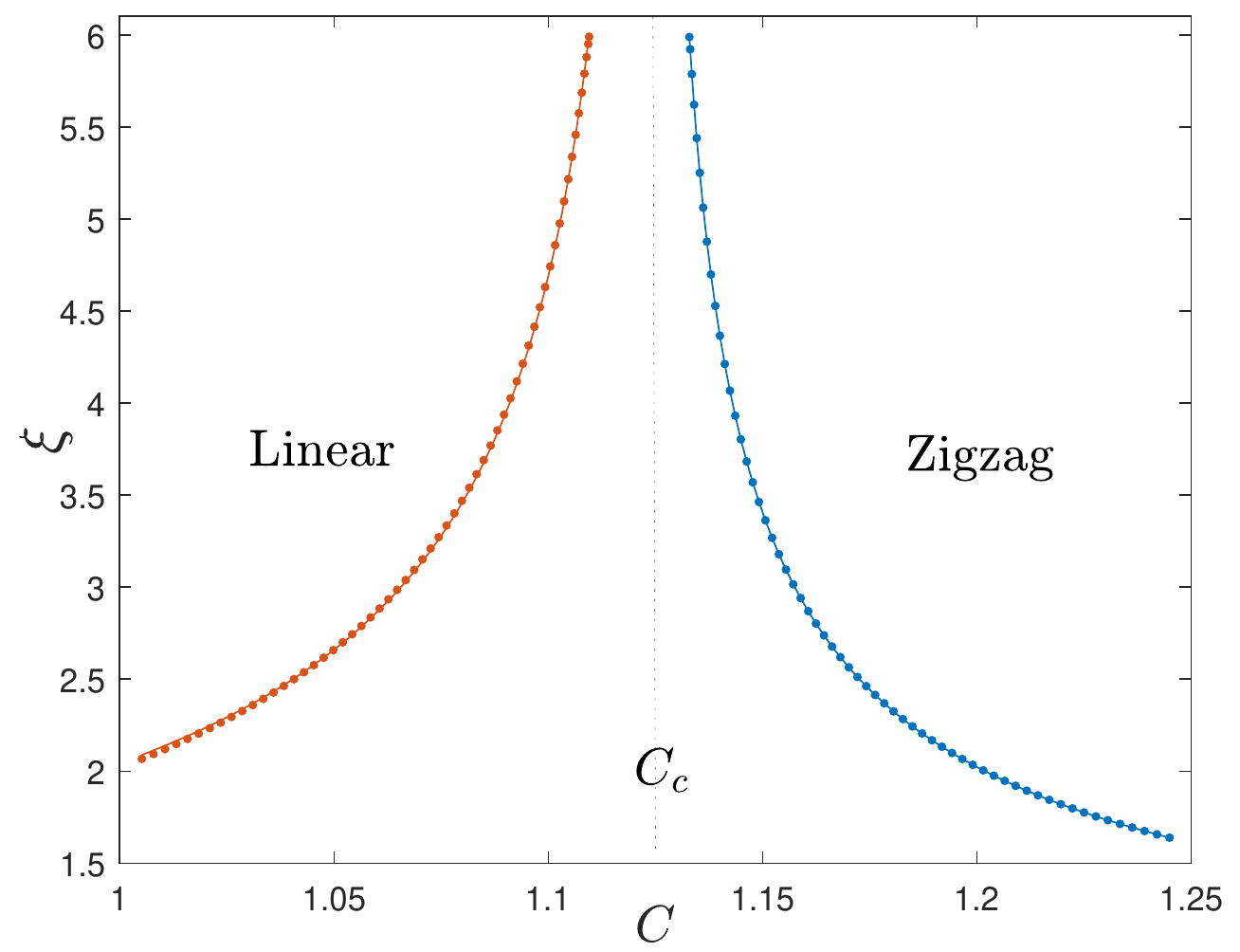}
\caption{The correlation length $\xi$ as a function of the coupling $C$. Blue dots represent numerical results, and the solid line is the fitted curve. The  red line fit is given by $(0.699\pm 0.003)(C_c-C)^{-0.516\pm 0.001}$ and the  blue line corresponds to $(0.579\pm 0.004)(C-C_c)^{-0.482\pm 0.002}$, where $C_c=1.125$ and $N=90$. Given the equilibrium scaling relation $\xi=\xi_{0}|\epsilon|^{-\nu}$, these fits are consistent with the mean-field value $\nu=1/2$. 
}\label{Fig7}
\end{figure}

In this appendix, we  calculate the QNMs of  the scalar fields in the holographic setting, by linearly perturbing their equations of motions in the RN-AdS metric. Specifically, we perturb
the scalar field $\psi_n(z)$ with a zigzag fluctuation $\delta\psi(z)  e^{ - i \omega t + i k n}$, where $\delta \psi$ is real. The linear equation for the fluctuation
reads
\begin{eqnarray}
& & \delta \psi (z) \left(-8 C z^2 \cos (k)+2 z \left(f'(z)-6 \psi_n (z)^2\right)-3 f(z)+3\right)\nonumber\\
& & +4z^2 \left(f(z) \delta \psi ''(z)+\delta \psi '(z) \left(f'(z)-2 i w\right)\right)=0. \label{qnm}
\end{eqnarray}
Note that  $\psi_n=0$ when $C< C_c$ corresponds to the linear state, while 
the non-trivial solution $\psi_n(z)=|\psi_1(z)|(-1)^{n}$ occurs when 
$C>C_c$ is the zigzag state.

Setting $\omega=0$, one can get a series QNMs of $k$ from the Eq.  \eqref{qnm}. Numerical results prove that all the real parts of QNM's $k$ are $\pi$, corresponding to the zigzag configuration. The correlation length $\xi=1/|{\rm Im}(k^*)|$ where $k^*$ is the lowest mode whose imaginary part is closest to the real axis. 
Similarly, the relaxation time $\tau$ can be obtained from Eq.  \eqref{qnm}  by setting $k=\pi$. One can get a series of modes of $\omega$. Then $\tau=1/|{\rm Im}(\omega^*)|$ where $\omega^*$ is the lowest mode in those QNMs.

\begin{figure}[t]
 \includegraphics[width=1.0\columnwidth,angle =0]{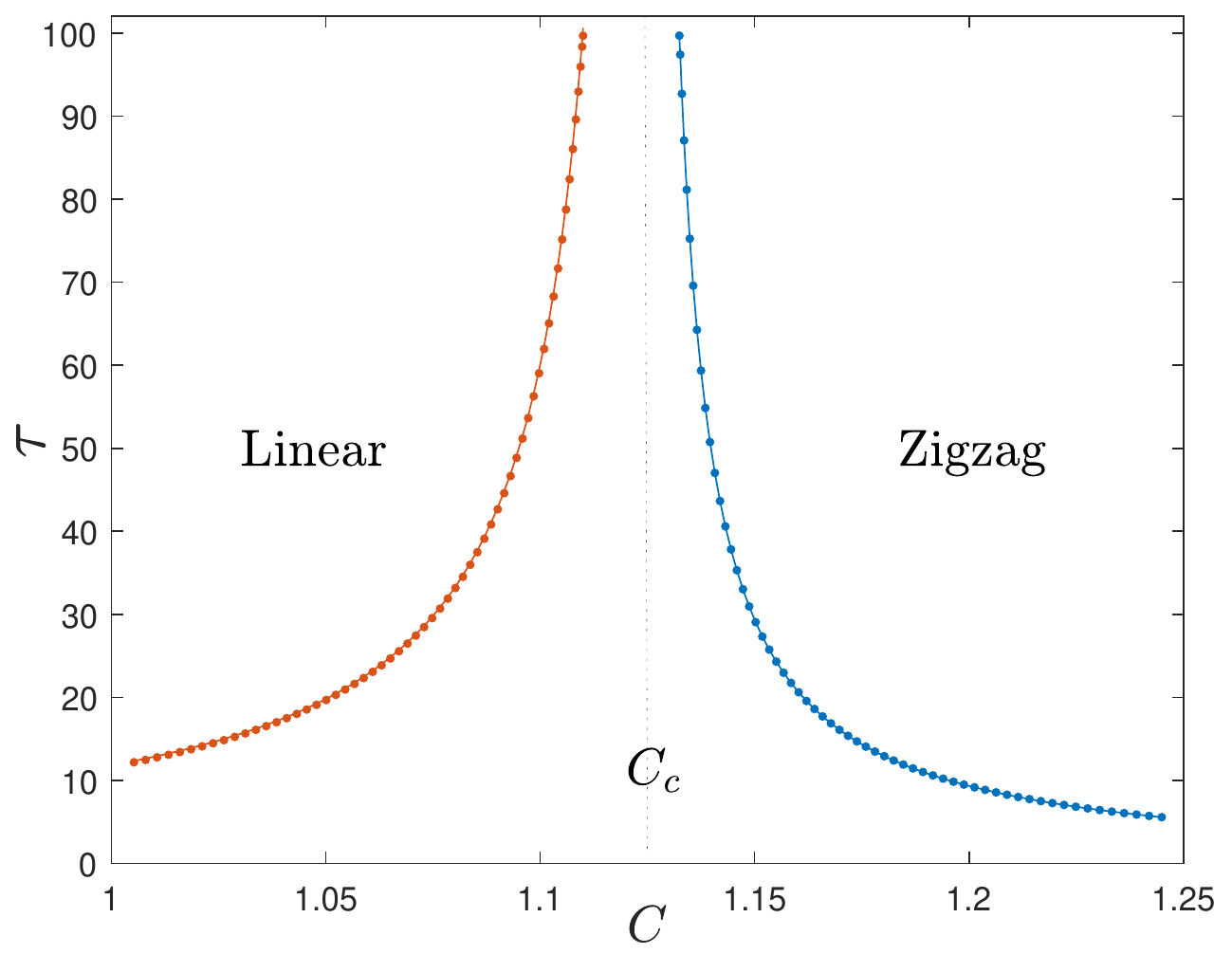}
\caption{The relaxation time $\tau$ as a function of the coupling $C$. The blue dots represent numerical results; the solid line is the fitted curve. The  red line represents the fit $(1.446\pm 0.004)(C_c-C)^{-1.008\pm 0.001}$ and the  blue line corresponds to $(0.679\pm 0.008)(C-C_c)^{-1.022\pm 0.003}$, where $C_c=1.125$ and $N=90$. Given the equilibrium scaling relation $\tau=\tau_{0}|\epsilon|^{-z\nu}$, these fits are thus consistent with the equilibrium critical exponents satisfying $z\nu=1$.
}\label{Fig8}
\end{figure}

The power-law divergence of the equilibrium correlation length $\xi$ as function of  the coupling $C$ is shown in Fig.  \ref{Fig7}. The power-law fit  reveals the value $\nu=0.516\pm 0.001$ in the high-symmetry phase  and $0.482\pm 0.002$ in the zigzag phase.  In spite of the finite-size effects, these values are in close agreement with  the  mean-field exponent $\nu=1/2$. Likewise, the critical slowing down associated with the divergence of the equilibrium relaxation time $\tau$ in the neighbourhood of the critical point  $C_c$ is shown in Fig.  \ref{Fig8}. In the linear  and zigzag phases, the fitted values of the power-law exponent are $1.008\pm 0.001$ and $1.022\pm 0.003$, respectively. These values are consistent with the relation $\nu z=1$. In conclusion, the fitted power-law
which indicates that $\nu=1/2$ and $z=2$, consistent with the mean-field theory. 

\bibliography{ref1}
\end{document}